\documentclass[prl,twocolumn,aps]{revtex4}

\usepackage{graphicx} 
\usepackage{color} 
\usepackage{epsfig}

\newcommand{\ket}[1]{\left|#1\right\rangle}

\begin{document}

\title{Quantum computing with nearest neighbor interactions and error rates over 1\%}

\author{David S. Wang, Austin G. Fowler, Lloyd C. L. Hollenberg}
\affiliation{Centre for Quantum Computer Technology, University of
Melbourne, Victoria, Australia}

\date{\today}

\begin{abstract}
Large-scale quantum computation will only be achieved if
experimentally implementable quantum error correction procedures are
devised that can tolerate experimentally achievable error rates. We
describe a quantum error correction procedure that requires only a
2-D square lattice of qubits that can interact with their nearest
neighbors, yet can tolerate quantum gate error rates over 1\%. The
precise maximum tolerable error rate depends on the error model, and
we calculate values in the range 1.1--1.4\% for various physically
reasonable models. Even the lowest value represents the highest
threshold error rate calculated to date in a geometrically
constrained setting, and a 50\% improvement over the previous
record.
\end{abstract}

\maketitle

Building a quantum computer is a daunting task. Engineering the
ability to interact nonlocal pairs of qubits is particularly
challenging. All existing quantum error correction (QEC) schemes
capable of tolerating error rates above 1\% assume the ability to
deterministically interact pairs of qubits separated by arbitrary
distances with no time or error rate penalty \cite{Knil04c, Fuji09,
Fuji10}. The most recent of these works estimates a threshold error
rate $p_{\rm th}$ of 5\% \cite{Fuji10}.

It is far more physically reasonable to assume a 2-D lattice of
qubits with only nearest neighbor interactions, proposed
realizations of which exist for ion traps \cite{Stoc08}, optical
lattices \cite{Jaks04}, superconducting qubits \cite{DiVi09},
optically addressed quantum dots \cite{VanM09, Jone10}, NV centers
in diamond \cite{Yao10} and many other systems. For such proposals,
the leading QEC scheme \cite{Raus07, Raus07d}, which is based on the
Kitaev surface code \cite{Brav98}, has been shown to possess a
$p_{\rm th}$ of 0.75\% \cite{Raus07, Fowl08, Wang09}. We increase
this to 1.1--1.4\%, depending on the error model, bringing the
threshold for geometrically constrained quantum computing above 1\%
for the first time. We achieve this by carefully using the given
error model to calculate approximate probabilities of different
error events and removing the need to initialize qubits.

To overview, we will begin by describing stabilizers and our
simplified quantum gate sequence, followed by a detailed discussion
of how probable different error events are and how this information
can be fed into the classical decoding algorithm. We then present
the results of detailed simulations, which apply two-qubit
depolarizing noise with probability $p_2$ after two-qubit quantum
gates, single-qubit depolarizing noise with probability $p_I$ after
identity gates, and make use of measurement gates that report and
project into the wrong eigenstate with probability $p_M$. In
addition to the standard error model with $p_2 = p_I = p_M = p$,
which we focus on, we simulate a balanced error model with $p_I =
4p_2/5$ and $p_M = 2p_I/3$, ensuring idle qubits have the same
probability of error as a single qubit involved in a two-qubit gate
and taking into account the fact that a measurement is only
sensitive to errors in one basis. We also simulate the case $p_I =
p_2/1000$ and $p_M = p_2/100$, modeling typical error ratios in an
ion trap.

\begin{figure}
\begin{center}
\resizebox{85mm}{!}{\includegraphics{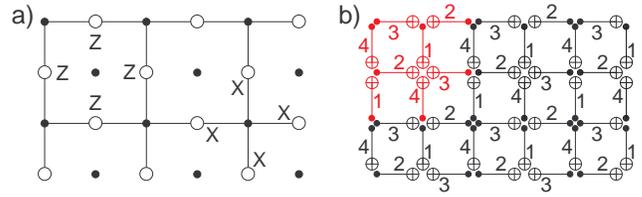}}
\end{center}
\caption{a) 2-D lattice of data qubits (circles) and syndrome qubits
(dots) and examples of the data qubit stabilizers. b) Sequence of
CNOTs permitting simultaneous measurement of all stabilizers.
Numbers indicate the relative timing of gates. The highlighted gates
can be tiled to fill the plain.}\label{transversal_sequence}
\end{figure}

\begin{figure}
\begin{center}
\resizebox{70mm}{!}{\includegraphics{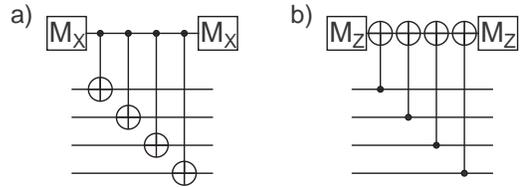}}
\end{center}
\caption{Circuit determining the sign of a stabilizer a) $XXXX$, b)
$ZZZZ$, without explicit initialization
gates.}\label{syndrome_measurements}
\end{figure}

A stabilizer \cite{Gott97} of a state $\ket{\Psi}$ is an operator
$S$ such that $S\ket{\Psi} = \ket{\Psi}$. An error $E$ that
anticommutes with $S$ can be detected as $SE\ket{\Psi} =
-ES\ket{\Psi} = -E\ket{\Psi}$. Examples of surface code stabilizers
\cite{Brav98} are shown in fig.~\ref{transversal_sequence}a.
Circuits measuring these stabilizers without explicit initialization
gates are shown in fig.~\ref{syndrome_measurements}. We assume
quantum nondemolition measurements, which have been experimentally
demonstrated using ion traps \cite{Burr10}, optical lattices
\cite{Krus10}, superconducting qubits \cite{Lupa06} and NV centers
in diamond \cite{Jele04} and theoretically proposed for optically
addressed quantum dots \cite{Bere06, Atat07}. The initial and final
measurements match when $+E$ is measured and differ when $-E$ is
measured. An appropriate sequence of two-qubit gates for measuring
all stabilizers across the lattice simultaneously is shown in
fig.~\ref{transversal_sequence}b. Data qubits execute identity gates
while the syndrome qubits are measured.

Repeatedly executing the gates of fig.~\ref{transversal_sequence}b,
along with appropriate syndrome qubit measurements and data qubit
identity gates, generates points in space and time where the
stabilizer changes sign, indicating local errors. Renormalization
techniques exist capable of processing perfect syndrome information
\cite{Ducl09, Ducl10}, however at present only the minimum weight
perfect matching algorithm \cite{Edmo65a, Edmo65b} can be used to
process the output of realistic quantum circuits.

The minimum weight perfect matching algorithm takes coordinates and
a measure of separation and matches pairs of coordinates such that
the total separation is a minimum. Chains of corrective operations
connecting matched pairs can then be applied. Prior work has
calculated the separation of two syndrome changes $s_1=(i_1, j_1,
t_1)$, $s_2=(i_2, j_2, t_2)$ using $d(s_1,
s_2)=|i_1-i_2|+|j_1-j_2|+|t_1-t_2|$ \cite{Raus07, Fowl08, Wang09},
however it was shown in \cite{Fowl10} that this is far from optimal
and leads to poor performance, particularly at low error rates. In
this work, we instead approximate the probability $P(s_1, s_2)$ of a
given pair of syndrome changes being connected by an error chain,
and set $d(s_1, s_2)=-\ln(P(s_1, s_2))$. This choice of $d(s_1,
s_2)$ is natural, accounting for the substantial performance
increase we observe.

\begin{figure}
\begin{center}
\resizebox{80mm}{!}{\includegraphics{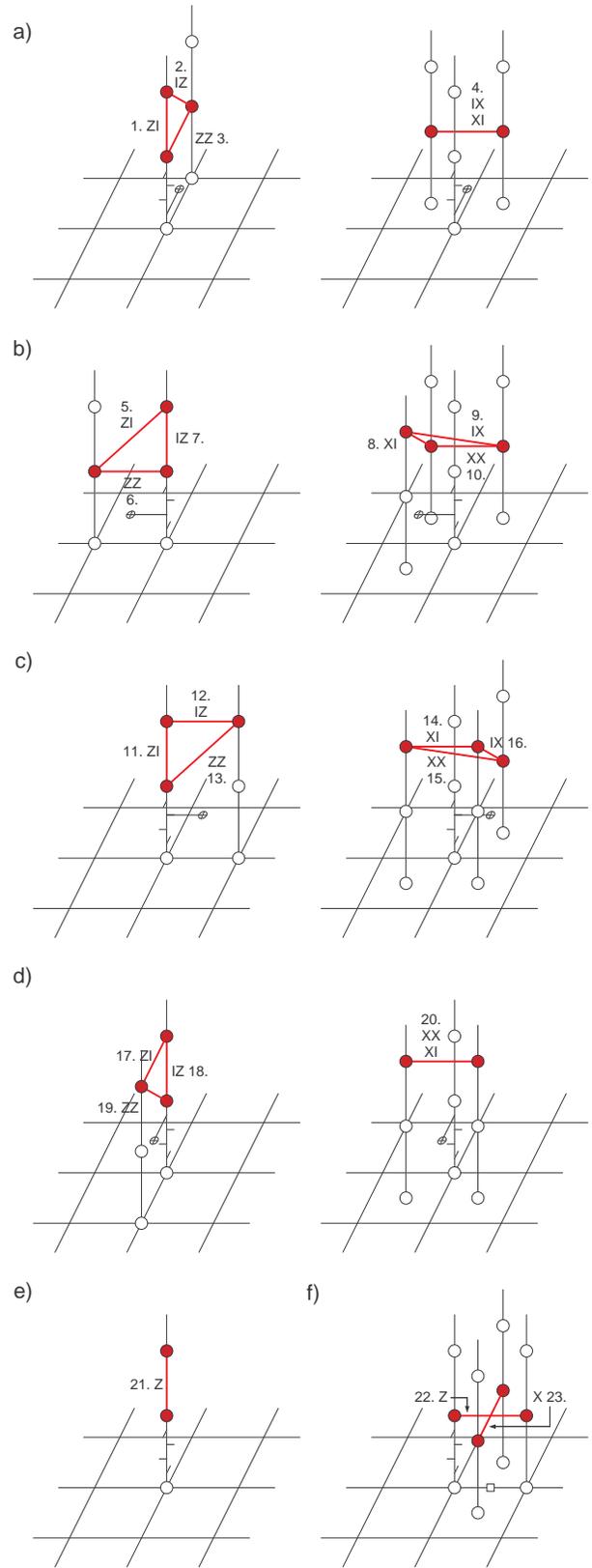}}
\end{center}
\caption{Syndrome changes resulting from a-d) specific two-qubit
errors on specific CNOTs, e) a syndrome qubit measurement error, f)
a data qubit memory error.}\label{links_from_gate2e-f}
\end{figure}

To calculate $P(s_1, s_2)$, we must study the effect gate errors.
Fig.~\ref{links_from_gate2e-f} shows all possible pairs of syndrome
changes resulting from all possible errors on all meaningfully
distinct gates. The CNOTs shown measure an $X$-stabilizer. The
effect of errors on the CNOTs used to measure a $Z$-stabilizer can
be obtained by interchanging $X$ and $Z$.

\begin{figure}
\begin{center}
\resizebox{80mm}{!}{\includegraphics{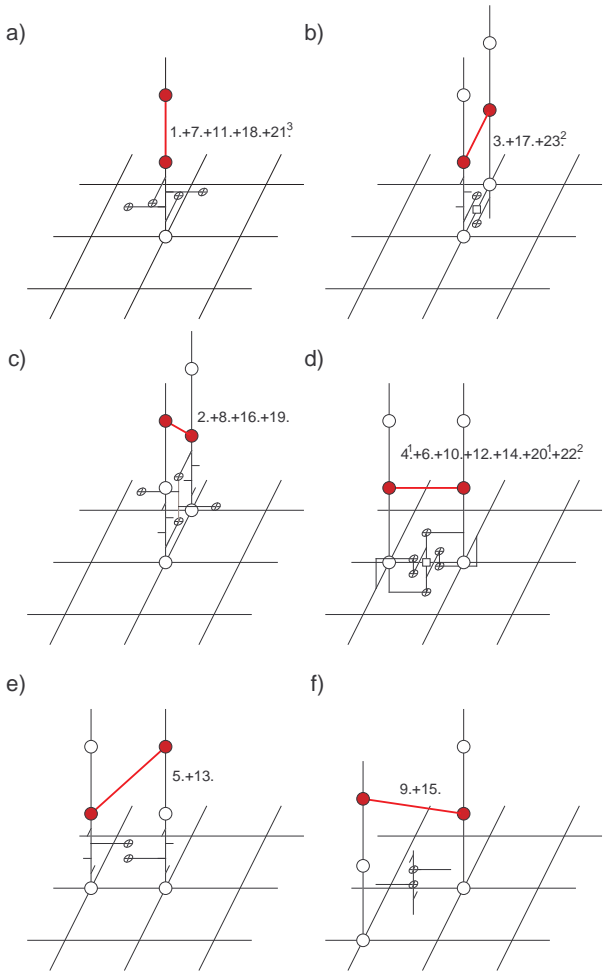}}
\end{center}
\caption{Numbered error processes from
fig.~\ref{links_from_gate2e-f} contributing to specific links.
Superscripts 1, 2 and 3 indicate errors occurring with probability
$8p_2/15$, $2p_I/3$ and $p_M$ respectively. All others occur with
probability $4p_2/15$.}\label{fig:link_from_gatesA-F}
\end{figure}

\begin{figure}
\begin{center}
\resizebox{55mm}{!}{\includegraphics{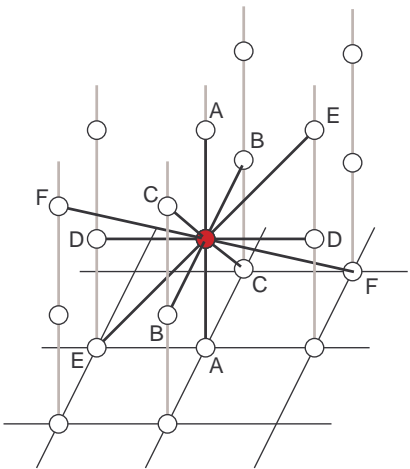}}
\end{center}
\caption{All possible links from a syndrome change to its neighbors.
Letters correspond to
figs.~\ref{fig:link_from_gatesA-F}a--f.}\label{fig:link_summary}
\end{figure}

Using fig.~\ref{links_from_gate2e-f},
fig.~\ref{fig:link_from_gatesA-F} was constructed, grouping gate
errors leading to specific pairs of syndrome changes. The
probability of an odd number of errors occurring in each group gives
the probability of the associated link. Using the standard error
model, the probability of the link shown in
fig.~\ref{fig:link_from_gatesA-F}a is
\begin{equation}
p_A = \frac{16p}{15}\left(1-\frac{4p}{15}\right)^3\left(1-p\right) +
p\left(1-\frac{4p}{15}\right)^4 + O(p^3).
\end{equation}
Defining similar probabilities $p_B$, $p_C$, $p_D$, $p_E$, $p_F$ for
fig.~\ref{fig:link_from_gatesA-F}b--f, fig.~\ref{fig:link_summary}
shows the $O(p)$ links from one syndrome to its neighbors. Some
straightforward modifications of the links and expressions are
required at the temporal and spatial boundaries.

\begin{figure}
\begin{center}
\resizebox{80mm}{!}{\includegraphics{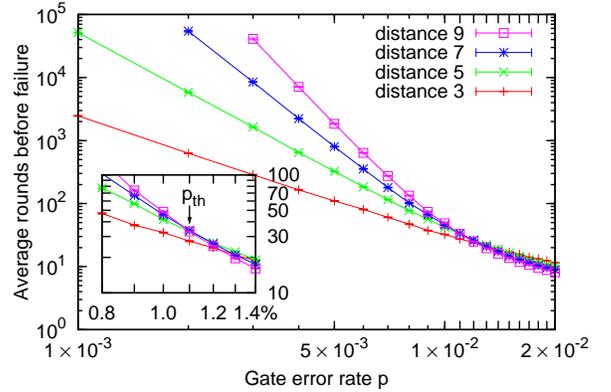}}
\end{center}
\caption{Average rounds of error correction before logical $X$
failure as a function of the gate error rate $p$ when using $d_{\rm
max}(s_1, s_2)$ and the standard error
model.}\label{fig:dtx_results}
\end{figure}

\begin{figure}
\begin{center}
\resizebox{80mm}{!}{\includegraphics{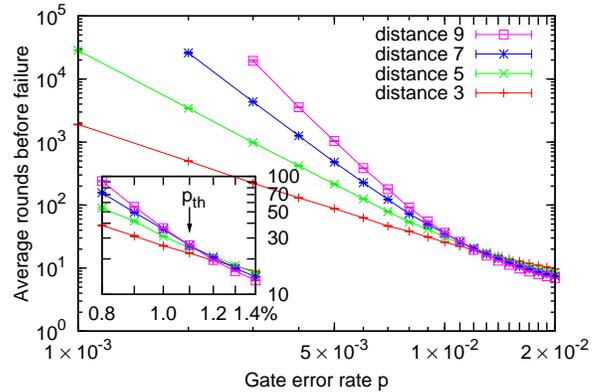}}
\end{center}
\caption{Average rounds of error correction before logical $Z$
failure as a function of the gate error rate $p$ when using $d_{\rm
max}(s_1, s_2)$ and the standard error
model.}\label{fig:dtz_results}
\end{figure}

The probability $P(s_1, s_2)$ that two syndrome changes are
connected is the sum of the probabilities of all connecting paths.
The probability of a given path is the product of the link
probabilities along the path. Several approximations of $P(s_1,
s_2)$ are worthy of study. The simplest approximation is to take a
single path of maximum probability $P_{\rm max}(s_1, s_2)$ and
define $d_{\rm max}(s_1, s_2)=-\ln(P_{\rm max}(s_1, s_2))$. We shall
see that this approximation is sufficient to substantially increase
$p_{\rm th}$, and that more accurate approximations do not lead to
further increase. The performance of surface code QEC using $d_{\rm
max}(s_1, s_2)$ and the standard error model is shown in
figs.~\ref{fig:dtx_results}--\ref{fig:dtz_results}.

\begin{figure}
\begin{center}
\resizebox{80mm}{!}{\includegraphics{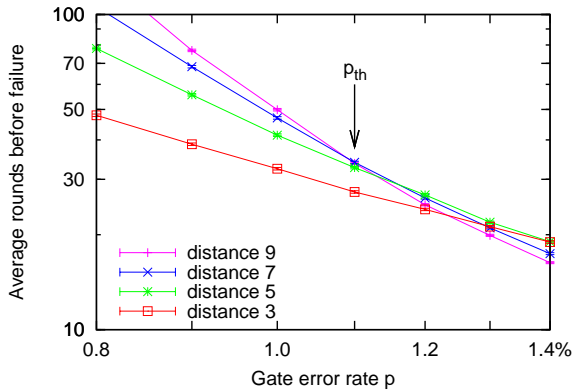}}
\end{center}
\caption{Average rounds of error correction before logical $X$
failure as a function of the gate error rate $p$ when using
$d_0(s_1, s_2)$ and the standard error model. $p_{\rm th}$ remains
1.1\%.}\label{fig:adv_results}
\end{figure}

Figs.~\ref{fig:dtx_results}--\ref{fig:dtz_results} give strong
evidence of $p_{\rm th}=1.1\%$. We have verified this by simulating
high distance codes at $p = 1.1\%$ and observing neither increase
nor decrease of the failure time. This is enormously encouraging,
and motivates one to better approximate $P(s_1, s_2)$ in an effort
to further increase $p_{\rm th}$. Additional accuracy can be
achieved by taking all shortest length paths (measured in links)
between $s_1$ and $s_2$ and calculating the sum of products of link
probabilities along each path. We shall define the resulting
distance measure as $d_0(s_1, s_2)$. We can define similar distance
measures $d_n(s_1, s_2)$ taking into account all minimum length $l$
paths and paths of length no greater than $l+n$. The performance of
surface code QEC around $p_{\rm th}$ using $d_0(s_1, s_2)$ is shown
in fig.~\ref{fig:adv_results}. It can be seen that $p_{\rm th}$
remains 1.1\%.

The fact that $d_0$ results in the same $p_{\rm th}$ as using a
single maximum probability path distance measure $d_{\rm max}$ can
be explained by noting that $d_0(s_1, s_2)$ only differs from
$d_{\rm max}(s_1, s_2)$ if $s_1$, $s_2$ are separated by at least
two links. Single link paths are unique minimum length paths
implying $d_0=d_{\rm max}$. The vast majority of error chains, even
for $p=p_{\rm th}$, are single links. We find that the modification
of the distance associated with the multiple paths of syndrome
changes separated by multiple links is in the region of 10--20\%.
Given such multiple link paths are not leading order contributors to
$p_{\rm th}$ in the first place, this relatively small weight change
does not result in an observable improvement of $p_{\rm th}$.

Higher order approximations $d_n(s_1, s_2)$ will also result in the
same $p_{\rm th}$ as the distance is hardly altered by increasing
$n$. To take a typical example, for $p=0.01$, $s_1 = (0,0,0)$, $s_2
= (1,-1,0)$ we obtain $d_0 = 6.91$ (6 paths), $d_1 = 6.86$ (30
paths) and $d_2 = 6.85$ (390 paths). The exponential increase of the
number of paths is well balanced by the exponential decrease of the
probability of these paths.

The balanced error model is a better model of all quantum gates
failing with equal probability than the standard error model, and
appropriate modification of the polynomials using $p_2 = p$, $p_I =
4p/5$ and $p_M = 8p/15$ leads to $p_{\rm th}=1.2\%$. The ion trap
error model, with $p_2 = p$, $p_I = p/1000$ and $p_M = p/100$ leads
to $p_{\rm th} = 1.4\%$. Arbitrary stochastic error models are
straightforward to analyze using our formalism.

To conclude, by performing a detailed study of the probability of
different pairs of syndrome changes and feeding the simplest
approximation of this information into the minimum weight perfect
matching algorithm, we have been able to raise the geometrically
constrained threshold error rate to 1.1--1.4\%, depending on the
exact error model, while maintaining computational efficiency. This
is the first time a geometrically constrained threshold error rate
has been observed over the 1\% level. There is the potential for
still further improvement by taking into account correlations
between $X$ and $Z$ errors, which shall be pursued in further work.

We acknowledge support from the Australian Research Council, the
Australian Government, and the US National Security Agency (NSA) and
the Army Research Office (ARO) under contract W911NF-08-1-0527.

\bibliography{../../../References}

\end{document}